# GPS IONOSPHERIC MAPPING AND TOMOGRAPHY: A CASE OF STUDY IN A GEOMAGNETIC STORM

*Shuanggen Jin[1*] and Rui Jin[1, 2]*

[1] Shanghai Astronomical Observatory, Chinese Academy of Sciences, Shanghai 200030, China
[2] Graduate University of the Chinese Academy of Sciences, Beijing 100049, China
Email: sgjin@shao.ac.cn; shuanggen.jin@gmail.com
Tel: 86-21-64386191-551; Fax: 86-21-64384618

## ABSTRACT

The ionosphere has been normally detected by traditional instruments, such as ionosonde, scatter radars, topside sounders onboard satellites and in situ rocket. However, most instruments are expensive and also restricted to either the bottomside ionosphere or the lower part of the topside ionosphere (usually lower than 800 km), such as ground based radar measurements. Nowadays, GPS satellites in high altitude orbits (~20,200 km) are capable of providing details on the structure of the entire ionosphere, even the plasmasphere. In this paper, a Regional Ionospheric Mapping and Tomography (RIMT) tool was developed, which can be used to retrieve 2-D TEC and 3-D ionospheric electron density profiles using ground-based or space-borne GPS measurements. Some results are presented from the RIMT tool using regional GPS networks in South Korea and validated using the independent ionosonde. GPS can provide time-varying ionospheric profiles and information at any specified grid related to ionospheric activities and states, including the electron density response at the F2-layer peak (the NmF2) during geomagnetic storms.

## 1. INTRODUCTION

The ionosphere is about 60-1000 km above the earth's surface, which is actually plasma of ionized gas of the upper atmosphere by solar radiation and high-energy particles from the Sun. The ionized electrons concentrations change with height above earth's surface, location, time of the day, season, and amount of solar activity. The total electron content (TEC) and electron density profiles are two key parameters in the ionosphere. Therefore, imaging the TEC and electron density profile is very crucial to determine the status of the ionospheric activities. In addition, the TEC in the Earth's ionosphere and plasmasphere is important for estimation and correction of propagation delays in satellite positioning, and predicting space weather and ionospheric disturbances due to geomagnetic storms and solar flares, etc.

In the past decades, different observing instruments have been developed and used to gather information on the ionosphere, such as ionosonde, scatter radars, topside sounders onboard satellites, in situ rocket and satellite observations and LEO (Low Earth Orbit) GPS occultation measurements. However, most instruments are expensive and also restricted to either the bottomside ionosphere or the lower part of the topside ionosphere (usually lower than 800 km), such as ground based radar measurements. Nowadays, GPS has become the most widely used tool for investigations of ionospheric irregularities due to the low-cost, all-weather, near real time, and high-temporal resolution (30 s) technique. Mapping TEC with GPS has been reported in the literatures and their application to space climate was successfully carried out by many countries' scientists [1]. In addition, most scientists used ground-based GPS data to monitor ionospheric irregularities based on a single layer model (SLM) of the ionosphere at the altitude of electron density peak (generally 350 km above the Earth). In fact, the single layer model ignored the vertical variation information of the ionosphere. Therefore, to better monitor the ionospheric activities in full dimensions, it is a new challenge to map 3-D ionosphere with GPS. Since Austen et al [2] first proposed the possibility of studying the ionosphere using satellite radio tomography, tomographic reconstruction techniques have been applied recently to ionospheric imaging.

In order to sufficiently and conveniently use GPS in the ionosphere by more uses, in this paper, we aim to develop a software tool, called Regional Ionospheric Mapping and Tomography software (RIMT), which can monitor 2-D TEC and map 3D ionospheric electron density distribution using GPS measurements. Some results on electron density response during the 20 November 2003 geomagnetic storm are presented from the RIMT tool using regional GPS networks in South Korea and validated using the independent ionosonde as well as discussed using the O/N2 ratio obtained from the GUVI instruments aboard the TIMED satellite.

## 2. GPS IONOSPHERIC MAPPING AND TOMOGRAPHY

### 2.1. 2-D ionospheric grid model

Usually the ionosphere consists of four layers, the D, E, F1 and F2 layers, and the peak ionospheric electron density locates at the F2 layer. According to the estimated vertical electron density distribution deriving from the Chapman profile, 90% electrons of the whole ionosphere are focusing from 100km to 800km. As the ionosphere is a dispersive medium, the codes and phases observations are delayed when the GPS signals propagate through it. The equations of GPS observations are expressed as follows [3]:

$$L_{k,j}^i = \lambda_k \phi_{k,j}^i = \rho_{0,j}^i - d_{ion,k,j}^i + d_{trop,j}^i + c(\tau^i - \tau_j) - \lambda_k (b_{k,j}^i + N_{k,j}^i) \quad (1)$$

$$P_{k,j}^i = \rho_{0,j}^i + d_{ion,k,j}^i + d_{trop,j}^i + c(\tau^i - \tau_j) + d_{q,k}^i + d_{q,k,j}^i + \varepsilon_j^i \quad (2)$$

where the subscript $k$ stands for the frequency, subscript $j$ is the ground-based GPS receiver number, superscript $i$ is the GPS satellites PRN, and other parameters are respectively:

$\rho_0$, the true distance between the receiver and satellite

$d_{ion}$, $d_{trop}$, the ionospheric and tropospheric delays

$c$, the speed of light in vacuum space

$\tau$, the satellite or receiver clock offset

$b$, the phase delay of satellite and receiver instrument bias

$d_q$, the codes delay of satellite and receiver instrument bias

$\lambda$, the carrier wavelength

$\phi$, the carrier phase observations

$N$, the ambiguity of the carrier phase

$\varepsilon$, other residuals such as multipath effect and noise

Using dual-frequency GPS observations, the following equations can be obtained as

$$L_4 = -40.3(\frac{1}{f_1^2} - \frac{1}{f_2^2}) STEC + B4 \quad (5)$$

$$P4 = 40.3(\frac{1}{f_1^2} - \frac{1}{f_2^2}) STEC + b4 \quad (6)$$

where $f$ is the frequency of GPS, STEC is the slant total electron content, $B4 = -\lambda_1(b_{1,j}^i + N_{1,j}^i) + \lambda_2(b_{2,j}^i + N_{2,j}^i)$ and $b4 = (dq_{1,j} - dq_{2,j}) + (dq_1^i - dq_2^i)$. STEC can be obtained from dual-frequency GPS carrier phase and code observations. The vertical total electron content (VTEC) are transferred from STEC through mapping function, i.e.,

$$VTEC = STEC / [1 - (\frac{R \cos(el)}{R + h})^2]^{-1/2} \quad (7)$$

where $R$ stands for the average radius of the earth, $h$ denotes for the single shell ionosphere attitude., and $el$ is the satellite elevating angle.

The 2-D ionospheric grid model using the measured VTEC deriving from GPS dual frequency observations can be established with the harmonic function of latitude and solar time or the method of distance-cost weighted. To get precise VTEC, $B_4$ can be obtained as $\sum_{i=1}^{N}(P_4 + L_4 - b_4)/N$, where $N$ is the number of samples for one arc of GPS observations, and the GPS satellite and receiver differential codes bias (DCB) are corrected. Since some stations are not the international GNSS service (IGS) stations, the DCB should be estimated from GPS code and phase observations. GPS satellites' differential codes bias (DCB) is very stable, which can be considered as a constant in one day. The daily mean of GPS satellites' DCB are available at the website (ftp://cddis.nasa.gov/pub/gps/products/ionex/), and ground GPS receivers' DCB should be estimated. In general, two strategies are used to estimate GPS receivers DCB [4]. One is based on one assumption that ionosphere is constant in a short time. In this method, STEC is a function with ionospheric pierce point coordinates in an Earth-Sun axis coordinate system. In this way, STEC parameters and the receiver DCB can be simultaneously estimated using the least-square principle. The other strategy is based on the method of minimization of VTEC standard deviation. The standard deviation is expressed as follows:

$$\sigma_{t,u} = \sum_{n=1}^{N_t} \sigma_u(n) \quad (8)$$

$$\sigma_u(n) = \sqrt{\frac{1}{M_t} \sum_{m=1}^{M_t} (VTEC_u^m(n) - \overline{VTEC_u(n)})^2} \quad (9)$$

here $\sigma_{t,u}$ is the standard deviation of VTEC data, $n$ stands for epoch number, $m$ is the satellite number, and subscript $u$ is the GPS ground station. Using the least square with a number of GPS epoch observations, the DCB value and VTEC can be estimated. And then 2-D ionospheric grid model is established from the measured VTEC deriving from GPS dual frequency observations with the harmonic function of latitude and solar time or the method of distance-cost weighted.

### 2.2. 3-D ionospheric tomography

Although 2-D ionospheric grid model is easily constructed from dual-frequency GPS measurements, it is lack of vertical ionospheric profile information. How to draw the 3-

D ionospheric information from ground GPS observations is a challenge. Based on the Computerized Ionospheric Tomography (CIT) introduced by Austen et al. [2], 3-D ionospheric model can be developed using the tomography technique. Firstly, the ionosphere is divided into a number of grid pixels with a small cell where the electron density is assumed to be constant (Figure 1).

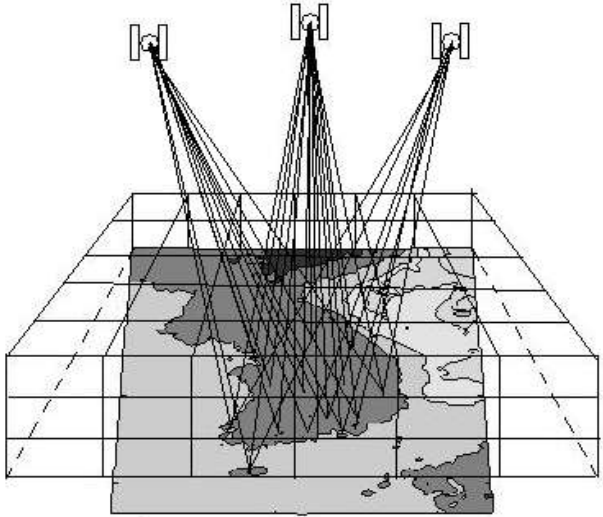

Figure 1 Geometric profile of mapping 3D ionosphere using ground-based GPS.

Secondly the STEC along the ray path $i$ can be approximately written as a finite sum over the pixels $j$ as follows:

$$STEC_i = \sum_{j=1}^{M} a_{ij} n_j \qquad (10)$$

Where aij denotes the length of the path-voxel intersections in the voxel $j$. $n_j$ is the electron density of the voxel $j$, which is related its longitude, latitude, height and time. Each set of STEC measurements from all observable satellites at consecutive epochs are combined into a linear expression:

$$Y = Ax + \varepsilon \qquad (11)$$

where Y is a column of m measurements of STEC, x is a column of n electron density in the targeted ionosphere region, and A is an m × n normal matrix with elements $a_{ij}$. The unknown electron densities $x$ can be estimated by the ionospheric tomographic reconstruction technique, e.g., multiplicative algebraic reconstruction technique (MART) as [5-7]

$$x_j^{k+1} = x_j^k \cdot \left( \frac{y_i}{\langle a_i, x^k \rangle} \right)^{\lambda_k a_{ij}} \qquad (12)$$

where $x_j$ is the $j$th resulted cell electron density in a column of n unknowns, $y_i$ is the $i$th STEC in a column of m measurements, $a_{ij}$ is the length of link $i$ that lies in cell j, $\lambda_k$ is the relaxation parameter at the $k$th iteration with $0 < \lambda_k < 1$, and the inner product of the vectors x and $a_i$ is the simulated STEC for the $i$th path. The electron density matrix x is therefore estimated iteratively by the ratio of the measured STEC and the simulated STEC with a relaxation parameter of $\lambda_k$ until the residual does not change. Here the latest IRI-2007 model (http://nssdcftp.gsfc.nasa.gov/models/ionospheric/iri/iri2007) is used as an initial guess for the reconstruction iteration. Thus, 3-D ionospheric tomography can be obtained from GPS observations with high sampling rate using the multiplicative algebraic reconstruction technique (MART).

## 3. IONOSPHERIC DISTURBANCES DURING THE GEOMAGNETIC STORM

In the following, the 3-D ionospheric disturbances during the large November 20th 2003 geomagnetic storm are investigated and analyzed using GPS data in South Korea. The geomagnetic storm Dst, Kp and AE indices on 20-21 November 2003 obtained from the World Data Center in Kyoto (http://swdcdb.kugi.kyoto-u.ac.jp/) showed a strong geomagnetic storm on November 20th, 2003 (Figure 2).

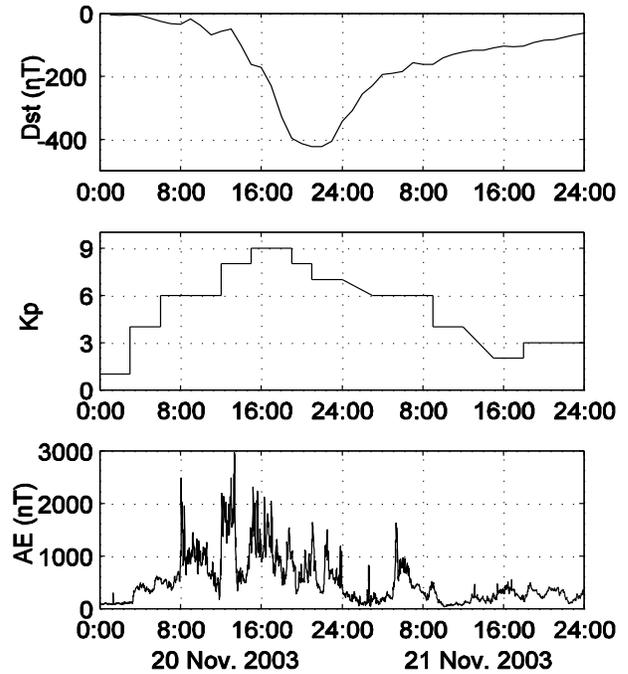

Figure 2. The geomagnetic storm index (Dst) (upper), Kp (middle) and AE indices (bottom) on 20-21 November 2003.

As the F2-layer peak electron density (NmF2) and its height (hmF2) are main parameters of the ionospheric electron density profile Ne (h), the behaviour of the ionospheric F2-layer to the storm was investigated over South Korea in terms of the NmF2 and hmF2. Here the peak density (NmF2) and its corresponding height (hmF2) are obtained from the ground-based GPS observations using the MART reconstructed tomography technique. The monthly median value of GPS reconstructed electron density profiles during the quiet days is regarded as the reference and the deviation of ionospheric NmF2 and hmF2 can reflect the ionospheric behaviors during the geomagnetic storm. It has shown that the GPS-derived NmF2 has a disturbance at 9:00 UT and then increases from 10:00 UT until 19:00 UT. The corresponding hmF2 also suddenly rises from 8:00 UT when the storm just started, and reaches the maximum height at about 16:00 UT with a maximum Kp value of 9, and then gradually descends until 21:00 UT (Figure 3), which are also supported by anther independent ionosonde measurement at Anyang station.

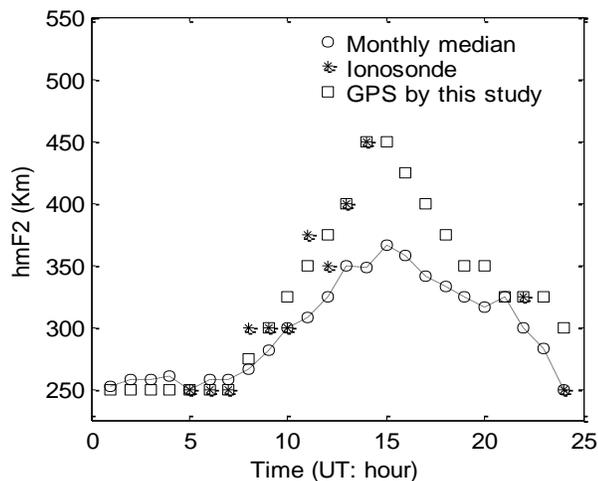

Figure 3. The hmF2 variations from monthly median, ionosonde and GPS observations on 20 November 2003.

Normally the increase/loss rate of F region electron density depends mainly on the molecular nitrogen concentration [N2] and atomic oxygen concentration [O] [8]. However, the O/N2 ratio obtained by the GUVI instrument on board the TIMED satellite doesn't show significant changes in South Korea where the increased NmF2 was observed, indicating that the increased NmF2 in South Korea is not caused by changes in neutral composition, and other possible non-chemical effects.

## 4. CONCLUSION

The two-dimensional and three-dimensional ionospheric information are obtained using ground GPS measurements over South Korea. The responses of the key ionospheric F2-layer parameters (NmF2 and hmF2) to the 20 November 2003 super storm have been studied using the GPS ionospheric reconstructed results. A strong increase of NmF2 during this storm has been found with corresponding significant hmF2 uplift, which is also supported by independent ionosonde observations at Anyang station. However, the O/N2 ratio from the GUVI instrument on board the TIMED satellite shows no significant change during this geomagnetic storm, indicating that the increase in NmF2 is not caused by changes in neutral composition, but is related to other possible non-chemical effects. It needs to be investigated in the future using more denser data.

## 5. ACKNOWLEDGMENT

The authors would like to thank those who made GPS observations available. This work was supported by the key program of Chinese Academy of Sciences (Grant No.KJCX2-EW-T03).